\documentclass[floatfix,twocolumn,aps,showpacs,showkeys,nofootinbib]{revtex4}

\usepackage{amssymb}
\usepackage{amsmath}
\usepackage{graphics}
\usepackage{epsfig}
\usepackage[dvips]{color}
\usepackage{bm}
\usepackage{color}

\newcommand{\be}{\begin{equation}}

\newcommand{\ee}{\end{equation}}
\newcommand{\bea}{\begin{eqnarray}}
\newcommand{\eea}{\end{eqnarray}}
\newcommand{\beas}{\begin{eqnarray*}}
\newcommand{\eeas}{\end{eqnarray*}}

\newcommand{\nn}{\nonumber\\}
\newcommand{\slsh}[1]{{\not \! #1}}

\begin{document}
\title{Inverse magnetic catalysis in the linear sigma model with quarks}
\author{Alejandro Ayala$^{1,3}$, M. Loewe$^{2,3}$, R. Zamora$^2$}
\affiliation{$^1$Instituto de Ciencias
  Nucleares, Universidad Nacional Aut\'onoma de M\'exico, Apartado
  Postal 70-543, M\'exico Distrito Federal 04510,
  Mexico.\\
  $^2$Instituto de F\1sica, Pontificia Universidad Cat\'olica de Chile,
  Casilla 306, Santiago 22, Chile.\\
  $^3$Centre for Theoretical and Mathematical Physics, and Department of Physics,
  University of Cape Town, Rondebosch 7700, South Africa}

\begin{abstract}

We compute the critical temperature for the chiral transition in the background of a magnetic field in the linear sigma model, including the quark contribution and the thermo-magnetic effects induced on the coupling constants at one loop level. We show that the critical temperature decreases as a function of the field strength. The effect of fermions on the critical temperature is small and the main effect on this observable comes from the charged pions. The findings support the idea that the inverse magnetic catalysis phenomenon receives a contribution due only to chiral symmetry effects independent of the deconfinement transition.

\end{abstract}



\maketitle

\section{Introduction}
\label{Introduction}
The properties of strongly interacting matter under the influence of magnetic fields exhibit intriguing properties. Recent lattice QCD results indicate that the transition temperature with 2 + 1 quark flavors, as measured from the behavior of the chiral condensate and susceptibility as well as from other thermodynamic observables such as longitudinal and transverse pressure, magnetization and energy and entropy densities, significantly decreases with increasing magnetic field~\cite{Fodor,Bali:2012zg,Bali2}. This result, which is obtained from simulations that get the physical pion mass, is nowadays widely accepted and differs from earlier lattice QCD calculations (which do not obtain the physical pion mass) that found an increasing critical temperature with the intensity of the field~\cite{Braguta, D'Elia:2011zu} as well as from several results obtained from model calculations~\cite{Loewe:2013coa, Agasian:2008tb, Fraga:2008qn, Mizher:2010zb, Andersen1}.

The above behavior has been dubbed {\it inverse magnetic catalsis} (though perhaps a more appropriate name could be {\it magnetic anticatalysis} given that the field acts against the formation of the condensate). Possible explanations include invoking a fermion paramagnetic contribution to the pressure with a sufficiently large magnetization~\cite{Noronha}, the competition between the valence and sea contributions at the phase transition~\cite{Bruckmann:2013oba} produced by a back reaction of the Polyakov loop which indirectly feels the magnetic field~\cite{Ferreira}, magnetic inhibition due to neutral meson fluctuations in a strong magnetic field~\cite{Fukushima} and the use of the functional renormalization group~\cite{Andersen2}. More recently, Refs.~\cite{Farias, Ferreira1} have postulated an ad hoc decreasing magnetic field and temperature dependent coupling, inspired by the QCD running of the coupling with energy, in the Nambu-Jona-Lasinio model. 

In a recent study~\cite{amlz} we have shown that the decrease of the coupling constant with increasing field strength can be obtained within a perturbative calculation in a model where charged scalars are subject to the effect of a constant magnetic field, the so called Abelian Higgs model. This behavior introduces in turn a dependence of the boson masses on the magnetic field and a decrease of the critical temperature for chiral symmetry breaking/restoration.  The essential ingredients for the calculation are the finite temperature effective potential in the presence of a magnetic field together with the proper handling of the plasma screening effects that have been recently consistently formulated for theories with spontaneous symmetry breaking~\cite{ahmrv}. The screening effects are included by means of the resummation of the ring diagrams which produce that for certain values of the model parameters and some temperatures the effective potential becomes an expansion containing a cubic term in the order parameter signaling that the description goes beyond the mean field approximation. 

The question that remains is to elucidate the simultaneous effects of quarks and bosons to explore the way they contribute to the inverse magnetic catalysis phenomenon. In this work we undertake such exploration in an effective QCD model. We use the linear sigma model with quarks to study the chiral symmetry breaking/restoration with magnetic fields. The work is organized as follows: In Sec.~\ref{efectivepotential} we recall the basic features of the linear sigma model with quarks in the presence of a magnetic field and write the effective potential at finite temperature including all of the model's degrees of freedom. In Sec.~\ref{III} we find the thermo-magnetic corrections to the boson self-coupling and the coupling between fermions and bosons. In Sec.~\ref{IV} we find the effects of the thermo-magnetic dependence of the couplings on the critical temperature for chiral symmetry restoration transition. We find that this critical temperature is a decreasing function of the magnetic field. We finally summarize and conclude in Sec.~\ref{conclusions}.

\section{Efective Potential}
\label{efectivepotential}

The model is given by the Lagrangian 
\begin{eqnarray}
   \mathcal{L}&=&\frac{1}{2}(\partial_\mu \sigma)^2  + \frac{1}{2}(D_\mu \vec{\pi})^2 + \frac{\mu^2}{2} (\sigma^2 + \vec{\pi}^2) - \frac{\lambda}{4} (\sigma^2 + \vec{\pi}^2)^2 \nonumber \\ 
   &+& i \bar{\psi} \gamma^\mu D_\mu \psi -g\bar{\psi} (\sigma + i \gamma_5 \vec{\tau} \cdot \vec{\pi} )\psi ,
\label{lagrangian}
\end{eqnarray}
where $\psi$ is an SU(2) isospin doublet, $\vec{\pi}=(\pi_1, \pi_2, \pi_3 )$ is an isospin triplet and $\sigma$ is an isospin singlet, with
\be
   D_{\mu}=\partial_{\mu}+iqA_{\mu},
\label{dcovariant}
\ee
is the covariant derivative. $A^\mu$ is the vector potential corresponding to an external magnetic field directed along the $\hat{z}$ axis,
\be
   A^\mu=\frac{B}{2}(0,-y,x,0),
\label{vecpot}
\ee
and $q$ is the particle's electric charge. $A^\mu$ satisfies the gauge condition $\partial_\mu A^\mu=0$. Since $A^3=0$, the gauge field only couples to the charged pion combinations, namely
\be
   \pi_\pm=\frac{1}{\sqrt{2}}\left(\pi_1\mp i\pi_2\right).
\ee
The neutral pion is taken as the third component of the pion isovector, $\pi^0=\pi_3$. The gauge field is taken as classical and thus we do not consider loops involving the propagator of the gauge field in internal lines. The squared mass parameter $\mu^2$ and the self-coupling $\lambda$ and $g$ are taken to be positive.

To allow for an spontaneous breaking of symmetry, we let the $\sigma$ field to develop a vacuum expectation value $v$
\be
   \sigma \rightarrow \sigma + v,
\label{shift}
\ee
which can later be taken as the order parameter of the theory.  After this shift, the Lagrangian can be rewritten as
\bea
   {\mathcal{L}} &=& -\frac{1}{2}[\sigma(\partial_{\mu}+iqA_{\mu})^{2}\sigma]-\frac{1}
   {2}\left(3\lambda v^{2}-\mu^{2} \right)\sigma^{2}\nn
   &-&\frac{1}{2}[\vec{\pi}(\partial_{\mu}+iqA_{\mu})^{2}\vec{\pi}]-\frac{1}{2}\left(\lambda v^{2}- \mu^2 \right)\vec{\pi}^{2}+\frac{\mu^{2}}{2}v^{2}\nn
  &-&\frac{\lambda}{4}v^{4} + i \bar{\psi} \gamma^\mu D_\mu \psi 
  -gv \bar{\psi}\psi + {\mathcal{L}}_{I}^b + {\mathcal{L}}_{I}^f,
  \label{lagranreal}
\eea
where ${\mathcal{L}}_{I}^b$ and  ${\mathcal{L}}_{I}^f$ are given by
\begin{eqnarray}
  {\mathcal{L}}_{I}^b&=&-\frac{\lambda}{4}\Big[(\sigma^2 + (\pi^0)^2)^2\nn 
  &+& 4\pi^+\pi^-(\sigma^2 + (\pi^0)^2 + \pi^+\pi^-)\Big],\nn
  {\mathcal{L}}_{I}^f&=&-g\bar{\psi} (\sigma + i \gamma_5 \vec{\tau} \cdot \vec{\pi} )\psi,
  \label{lagranint}
\end{eqnarray}
and represent the Lagrangian describing the interactions among the fields $\sigma$, $\vec{\pi}$ and $\psi$, after symmetry breaking. From Eq.~(\ref{lagranreal}) we see that the $\sigma$, the three pions and the quarks have masses given by
\bea
  m^{2}_{\sigma}&=&3  \lambda v^{2}-\mu^{2},\nn
  m^{2}_{\pi}&=&\lambda v^{2}-\mu^{2}, \nn
  m_{f}&=& gv,
\label{masses}
\eea
respectively.

Using Schwinger's proper-time method, the expression for the one-loop effective potential for one boson field with squared mass $m_b^2$ and absolute value of its charge $q_b$ at finite temperature $T$ in the presence of a constant magnetic field can be written as
\bea
  V_b^{(1)} &=& \frac{T}{2}\sum_n\int dm_b^2\int\frac{d^3k}{(2\pi)^3}\int_0^\infty
   \frac{ds}{\cosh (q_bBs)}\nn
   &\times&e^{-s(\omega_n^2+k_3^2 + k_\perp^2\frac{\tanh (q_bBs)}{q_bBs} + m_b^2)},
   \label{boson1}
\eea
where $\omega_n=2n\pi T$ are boson Matsubara frequencies.
Similarly, the expression for the one-loop effective potential for one fermion field with mass $m_f$ and absolute value of its charge $q_f$ at finite temperature $T$ in the presence of a constant magnetic field can be written as
\bea
  V_f^{(1)}&=& -\sum_{r=\pm 1}T\sum_n\int dm_f^2\int\frac{d^3k}{(2\pi)^3}\int_0^\infty
   \frac{ds}{\cosh (q_fBs)}\nn
   &\times&e^{-s(\tilde{\omega}_n^2+k_3^2 + k_\perp^2\frac{\tanh (q_fBs)}{q_fBs} + m_f^2 +r q_fB)},
   \label{fermion1}
\eea
where $\tilde{\omega}_n=(2n+1)\pi T$ are fermion Matsubara frequencies. The sum over the index $r$ corresponds to the two possible spin orientations along the magnetic field direction. 

\begin{figure}[h]
\begin{center}
\includegraphics[scale=0.6]{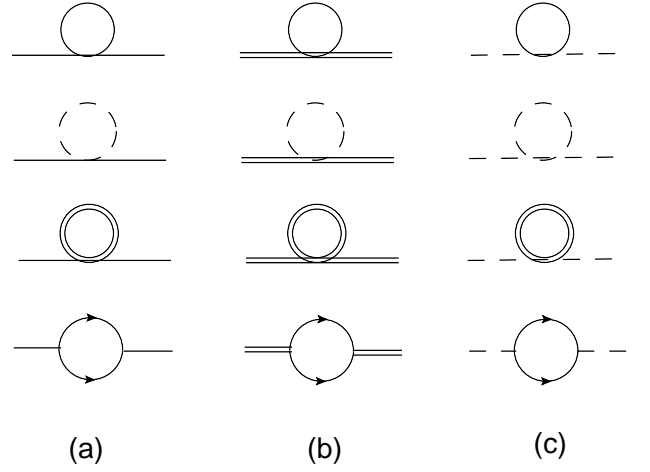}
\end{center}
\caption{Feynman diagrams contributing to the one loop bosons' self-energies. The dashed line denotes the charged pion, the continuous line is the sigma, the double line represents the neutral pion and the continuous line with arrows represents the fermions.}
\label{fig1}
\end{figure}

Including the $v$-independent terms, choosing the renormalization scale as $\tilde{\mu}=e^{-1/2}\mu$ and after mass and charge renormalization, it has been shown in Ref.~\cite{ahmrv} that the thermo-magnetic effective potential in the small to intermediate field regime, in a high temperature expansion can be written as
\bea
   V^{({\mbox{\small{eff}}})}&=&
   -\frac{\mu^2}{2}v^2 + \frac{\lambda}{4}v^4\nn
   &+&\sum_{i=\sigma,\pi^0}\left\{\frac{m_i^4}{64\pi^2}\left[ \ln\left(\frac{(4\pi T)^2}{2\mu^2}\right) 
   -2\gamma_E +1\right]
   \right.\nn
   &-&\left. \frac{\pi^2T^4}{90} + \frac{m_i^2T^2}{24}  - \frac{T}{12 \pi}(m_i^2 + \Pi)^{3/2} \right\} \nn
   &+&\sum_{i=\pi_+,\pi_-}\left\{\frac{m_i^4}{64\pi^2}\left[ \ln\left(\frac{(4\pi T)^2}{2\mu^2}\right) 
   -2\gamma_E +1\right]
   \right.\nn
   &-&\frac{\pi^2T^4}{90} + \frac{m_i^2T^2}{24}\nn
   &+&\frac{T(2qB)^{3/2}}{8\pi}\zeta\left(-\frac{1}{2},\frac{1}{2}+\frac{m_i^2+\Pi}{2qB}\right)\nn
   &-&\frac{(qB)^2}{192\pi^2}\left[ \ln\left(\frac{(4\pi T)^2}{2\mu^2}\right)
   - 2\gamma_E + 1\right.\nn
   &+&\left.\left. 
   \zeta (3)\left(\frac{m_i}{2\pi T}\right)^2 - \frac{3}{4}\zeta (5) \left(\frac{m_i}{2\pi T}\right)^4   
   \right]\right\} \nn 
   &-& \sum_{f=u,d}\left\{\frac{m_f^4}{16\pi^2}\left[ \ln\left(\frac{(\pi T)^2}{2\mu^2}\right) 
   -2\gamma_E +1\right] \right. \nn
   &+&   \frac{7 \pi^2 T^4}{180} - \frac{m_f^2 T^2}{12}   \nn
   &+&  \left. \frac{(q_fB)^2}{24\pi^2}\left[ \ln\left(\frac{(\pi T)^2}{2\mu^2}\right) - 2\gamma_E + 1 \right] \right\},
   \label{Veff-mid}
\eea
where $q$ is the absolute value of the charged pions' charge ($q=1$),  $q_u=2/3$, $q_d=1/3$ are the absolute values of the $u$ and $d$ quarks, respectively and $\gamma_E$ is Euler's gamma. Though we take the quark masses as equal, the notation emphasizes that the effective potential is evaluated accounting for the different quark charges. We have introduced the leading temperature plasma screening effects for the boson's mass squared, encoded in the boson's self-energy $\Pi$. For the Hurwitz zeta function $\zeta(-1/2,z)$ in Eq.~(\ref{Veff-mid}) to be real, we need that 
\bea
   -\mu^2 + \Pi > qB,
\label{requirealso}
\eea
condition that comes from requiring that the second argument of the Hurwitz zeta function satisfies $z>0$, even for the lowest value of $m_b^2$ which is obtained for $v=0$. Furthermore, for the large $T$ expansion to be valid, we also require that
\bea
   qB/T^2 <1.
\label{otherrequirement}
\eea
The diagrams representing the bosons' self-energies are depicted in Fig.~\ref{fig1}. Each column corresponds to the diagrams contributing to the self-energy of a given boson. The total self-energy for any boson is identical to the other's and thus we concentrate on computing the diagrams in column (a). 
The contribution from the individual diagrams require of the expressions [hereby capital letters are used to denote four-momenta in Eucledian space, {e.g.} $K\equiv (\omega_n,{\mbox{\bf{k}}})$] 
\bea
   \Pi_{a1}(m_{\sigma}^2)&=&\lambda T \sum_n\int\frac{d^3k}{(2\pi)^3}D(K;m_{\sigma}^2),\nn 
   \Pi_{a2}(m_{\pi^0}^2)&=&\lambda T \sum_n\int\frac{d^3k}{(2\pi)^3}D(K;m_{\pi^0}^2)  \nn
   \Pi_{a3}(m_{\pi^\pm}^2)&=&\lambda T \sum_n\int\frac{d^3k}{(2\pi)^3}D_B(K;m_{\pi^\pm}^2),\nn
   \Pi_{a4}(P;m_f)&=&-N_f g^2T\sum_n\int\frac{d^3k}{(2\pi)^3}\nn
   &\times&{\mbox{Tr}}\ S_B(K;m_f)S_B(P-K;m_f),
\label{general1}
\eea
where $N_f$ is the number of fermions and the corresponding propagators are given by
\bea
   D(K;m_i^2)&=&\frac{1}{K^2+m_i^2},\nn
   D_B(K;m_i^2)&=&\int_0^\infty ds \frac{e^{-s(\omega_n^2+k_3^2+k_\perp^2
   \frac{\tanh (qBs)}{qBs} + m_i^2)}}{\cosh (qBs)},\nn
   S_B(K;m_f)&=&\int_0^\infty ds \frac{e^{-s(\widetilde{\omega}_n^2+k_3^2+k_\perp^2
   \frac{\tanh (q_fBs)}{q_fBs} + m_f^2)}}{\cosh (q_fBs)}\nn
   &\times&\!\!\!\left[(\cosh (q_fBs) -i \gamma_1 \gamma_2 \sinh (q_fBs)) \right. \nn 
   &\times& \left. (m_f-\slsh{k_{\|}}) - \frac{\slsh{k_\bot}}{\cosh(q_fBs)} \right],
\label{Schwinger}
\eea
and for the charged particle propagators we have used Schwinger's proper-time representation. For the computation of $\Pi_{a3}$ we work in the {\it infrared limit}, namely, $P=(0,{\mbox{\bf{p}}}\rightarrow 0)$ and with the hierarchy of scales $qB,m_i^2< T^2$. It has been shown~\cite{reloaded} that this limit can be formally implemented by straightforward setting $P=0$ in the third of Eqs.~(\ref{general1}). The leading contribution at high temperature from each of these diagram is
\bea
   \Pi_{a1}(m_\sigma^2)&=& \lambda \frac{T^2}{12}  \nn
   \Pi_{a2}(m_{\pi^0}^2)&=& \lambda \frac{T^2}{12}  \nn
   \Pi_{a3}(m_{\pi^\pm}^2)&=& \lambda \frac{T^2}{12}  \nn
   \Pi_{a4}(0;m_f)&=&N_f g^2\frac{T^2}{6},
\eea
and therefore, considering the permutation factors, the total self-energy is given by 
\bea
   \Pi&=& 3 \Pi_{a1}(m_\sigma^2) + \Pi_{a2}(m_{\pi^0}^2)\nn
       &+& 2\Pi_{a3}(m_{\pi^\pm}^2) + \Pi_{a4}(0;m_f)\nn
       &=&\lambda \frac{T^2}{2} + N_fg^2\frac{T^2}{6}.
\label{self}
\eea

\section{One loop thermo-magnetic couplings}\label{III}

\begin{widetext}
\begin{figure*}[ht]
\begin{center}
\includegraphics[scale=1]{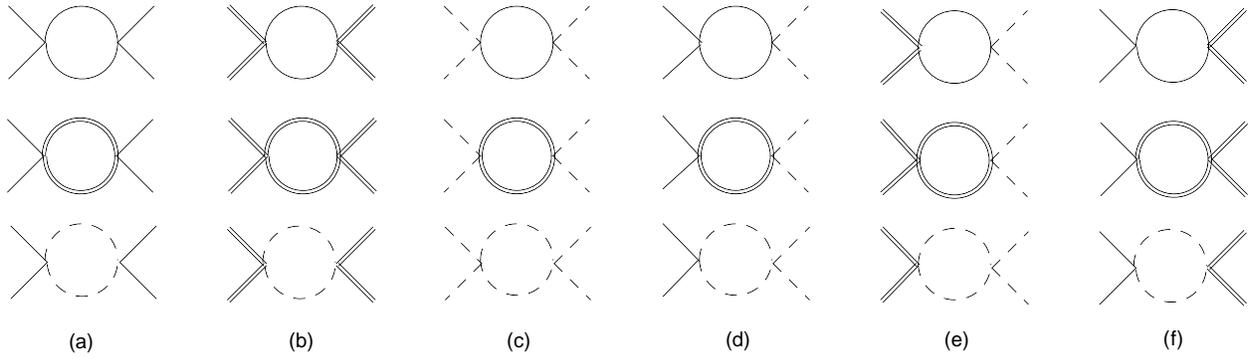}
\end{center}
\caption{One-loop Feynman diagrams that contribute to the thermal and magnetic correction to the coupling $\lambda$. The dashed line denotes the charged pion, the continuous line is the sigma and the double line represents
the neutral pion.}
\label{fig2}
\end{figure*}
\end{widetext}

Let us now compute the one-loop correction to the coupling $\lambda$, including thermal and magnetic effects. Figure~\ref{fig2} shows the Feynman diagrams that contribute to this correction. Columns (a), (b), (c), (d), (e) and (f) contribute to the correction to the $\sigma^4$, $(\pi^0)^4$, $(\pi^+)^2(\pi^-)^2$, $\sigma^2\pi^+\pi^-$, $(\pi^0)^2\pi^+\pi^-$ and  $\sigma^2(\pi^0)^2$ terms of the interaction Lagrangian in Eq.~(\ref{lagranint}), respectively. Since each of these corrections lead to the same result, we concentrate on the diagrams in column (a). Each of the three diagrams involves two propagators of the same boson. For the first two diagrams the intermediate bosons are neutral and for the third one the intermediate bosons are charged. Therefore the expression for the diagrams can be obtained from 
\bea
   I(P_i;m_i^2)&=&T\sum_n\int\frac{d^3k}{(2\pi)^3}D(P_i-K)D(K),\nn
   J(P_i;m_i^2)&=&T\sum_n\int\frac{d^3k}{(2\pi)^3}D_B(P_i-K)D_B(K),\nn
\label{general}
\eea
where $P_i$ is the total incoming four-momentum and $D$ and $D_B$ are the boson propagators defined in Eqs.~(\ref{Schwinger}). \\

\begin{figure}[t]
\vspace{0.4cm}
\begin{center}
\includegraphics[scale=0.49]{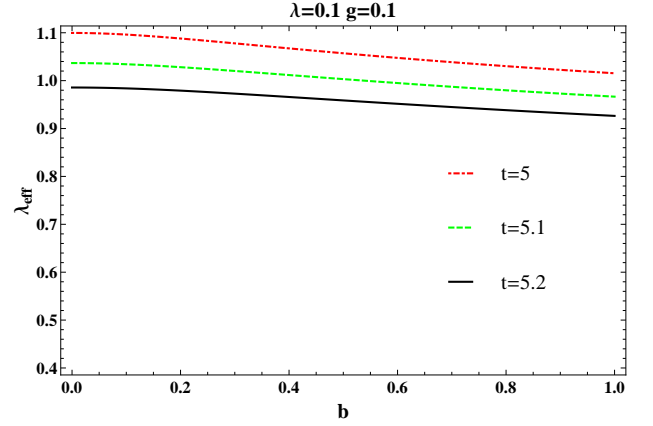}
\end{center}
\caption{Color on-line. $\lambda_{\mbox{\small{eff}}}(v=0)$ as a function of $b=qB/\mu^2$ for three different values of $t=T/\mu$.}
\label{fig3}
\end{figure}

Once again we work in the {\it infrared limit}, namely, $P_i=(0,{\mbox{\bf{p}}}\rightarrow 0)$ and with the hierarchy of scales where $qB,m_i^2< T^2$. It is known that in order to properly implement this hierarchy~\cite{alhmrv}, it is necessary to separate the contribution from the Matsubara zero mode from the rest of the modes in Eq.~(\ref{general}). We therefore write
\bea
   J(0;m_i^2)&\equiv& J_{n=0}(0;m_i^2) + J_{n\neq 0}(0;m_i^2)\nn
   J_{n\neq 0}(0;m_i^2)&=&T\sum_{n\neq 0}\int\frac{d^3k}{(2\pi)^3}
   D_B(\omega_n,{\mbox{\bf{k}}})D_B(\omega_n,{\mbox{\bf{k}}})\nn
   J_{n=0}(0;m_i^2)&=&T\int\frac{d^3k}{(2\pi)^3}D_B({\mbox{\bf{k}}})D_B({\mbox{\bf{k}}})
\label{separate}
\eea
$J_{n=0}$ is straightforward computed with the result
\bea
   J_{n=0}(0;m_i^2)=\frac{T}{16\pi}\frac{1}{(2qB)^{1/2}}  \zeta \left(\frac{3}{2}, \frac{1}{2} + \frac{m^2_{i}+\Pi}{2qB}\right),
\label{n=0}
\eea 
where we have also included the plasma screening effects for the boson's mass squared~\cite{ahmrv}. The contribution from the rest of the modes is performed by resorting to the weak field limit of the boson propagator~\cite{mexicanos}
\bea
   D_B(\omega_{n\neq 0},{\mbox{\bf{k}}})&=&\frac{1}{\omega_n^2+{\mbox{\bf{k}}}^2+m_i^2}
   \left[1-\frac{(qB)^2}{(\omega_n^2+{\mbox{\bf{k}}}^2+m_i^2)^2}\right.\nn
   &+&\left.\frac{2(qB)^2\ k_\perp^2}{(\omega_n^2+{\mbox{\bf{k}}}^2+m_i^2)^3}
   \right].
\label{propnneq0}
\eea
The sum and integrals in $J_{n\neq 0}$ are performed by means of the Mellin summation technique~\cite{Bedingham} with the result
\bea
   J_{n\neq 0}(0;m_i^2)&=&-\frac{1}{16 \pi^2} \Big[\ln\left( \frac{(4\pi T)^2}{2\mu^2} \right) +1 - 2\gamma_E\nn
   &+&\zeta(3) \left(\frac{\sqrt{m_i^2+\Pi}}{2 \pi T}\right)^2   \Big]\nn
   &-&\frac{(q B)^2 }{1024 \pi^6 T^4} \zeta(5).
\label{nneq0}
\eea
We have also included the plasma screening effects for the boson's mass squared and have carried out the mass renormalization introducing a counter term $\delta m^2=-1/\epsilon+\gamma_E-\ln(2\pi)$. The function $J(0;m_i^2)$ is therefore explicitly obtained by adding up Eqs.~(\ref{n=0}) and~(\ref{nneq0}). The function $I(0;m_i^2)$ is obtained as the limit when $qB\rightarrow 0$ of the function function $J(0;m_i^2)$, with the result
\bea
   I(0;m_i^2)&=&\frac{T}{8\pi}\frac{1}{(m_i^2 + \Pi)^{1/2}}\nn
   &-&\frac{1}{16 \pi^2} \Big[\ln\left( \frac{(4\pi T)^2}{2\mu^2} \right) +1 - 2\gamma_E\nn
   &+&\zeta(3) \left(\frac{\sqrt{m_i^2+\Pi}}{2 \pi T}\right)^2   \Big].
\label{nneq02}
\eea
Considering the permutation factors and the contribution from the $s$, $t$ and $u$-channels, the correction to the self-coupling $\lambda$ to one-loop order is given by
\bea
   \lambda_{\mbox{\small{eff}}}=\lambda
   \left[1 + 24\lambda\left( 9I(0;m_\sigma^2) + I(0;m_\pi^2) + 4 J(0;m_\pi^2) \right)
   \right].\nn
\label{lambdaeff}
\eea
Note that $\lambda_{\mbox{\small{eff}}}$ depends on $v$ through the dependence on the boson masses. Let us furthermore take the approximation where we evaluate $\lambda_{\mbox{\small{eff}}}$ at $v=0$. The rationale is that we are pursuing the effect on the critical temperature which is the temperature where the curvature of the effective potential at $v=0$ vanishes. Figure~\ref{fig3} shows the behavior of $\lambda_{\mbox{\small{eff}}}(v=0)$ as a function of $b=qB/\mu^2$ for three different values of $t=T/\mu$.  Note that in all cases the effective coupling is a decreasing function of the magnetic field strength.
\begin{figure}[t]
\begin{center}
\includegraphics[scale=0.5]{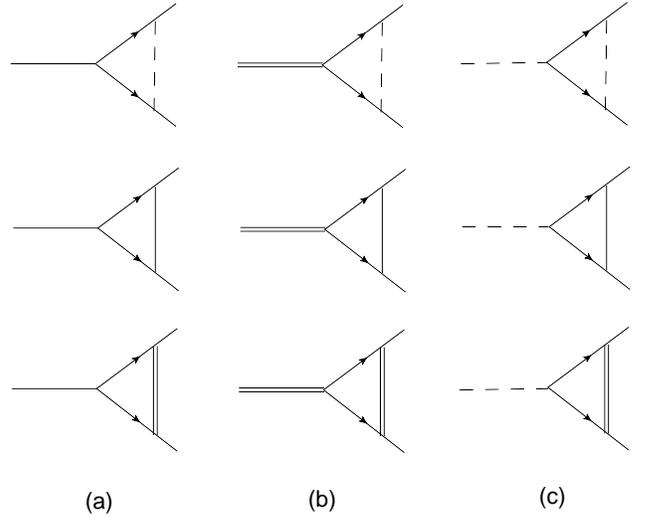}
\end{center}
\caption{One-loop Feynman diagrams that contribute to the thermal and magnetic correction to the coupling $g$. The dashed line denotes the charged pion, the continuous line is the sigma, the double line represents the neutral pion and the continuous line with arrows represents the quarks.}
\label{fig4}
\end{figure}
We now turn to the calculation of the thermo-magnetic correction of the coupling $g$. Figure~\ref{fig4} shows the Feynman diagrams that contribute to this correction. We are interested in computing an effective value for this coupling, $g_{\mbox{\small{eff}}}$, also for $v = 0$, in the same manner we did for $\lambda_{\mbox{\small{eff}}}$. Columns (a), (b) and (c) contribute to the correction to the quark-$\sigma$, quark-$\pi^0$ and quark-$\pi^\pm$ terms of the interaction Lagrangian of Eq.~(\ref{lagranint}), respectively. Since each of these corrections lead to the same result, we concentrate on the diagrams in column (a). Note that from Eq.~(\ref{masses}), for $v = 0$ the masses of $\sigma$ and $\pi^0$ become degenerate and since the middle and bottom diagrams contribute with opposite sign, they cancel [incidentally, this also happens with the two bottom diagrams in columns (b) and (c)].  We thus concentrate on the top diagram in column (a). The expression for this diagram is written as
\bea
   L(P_i;v) &=& 2 T g^3 \sum_n\int\frac{d^3k}{(2\pi)^3}S_B(K;m_u)\nn 
   &\times& S_B(P_i-K;m_d)D_B(P_i-K;m_{\pi}),\nn 
\label{general2}
\eea
\begin{figure}[b]
\begin{center}
\includegraphics[scale=0.48]{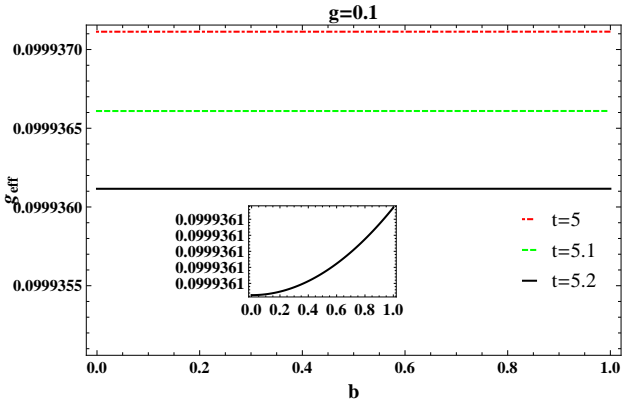}
\end{center}
\caption{Color on-line. $g_{\mbox{\small{eff}}}(v=0)$ as a function of $b=qB/\mu^2$ for three different values of $t=T/\mu$. $g_{\mbox{\small{eff}}}$ is a growing function of the magnetic field strength, though the growing is rather mild, this can be better appreciated in the inset.}
\label{fig5}
\end{figure}
\begin{figure}[h!]
\begin{center}
\includegraphics[scale=0.5]{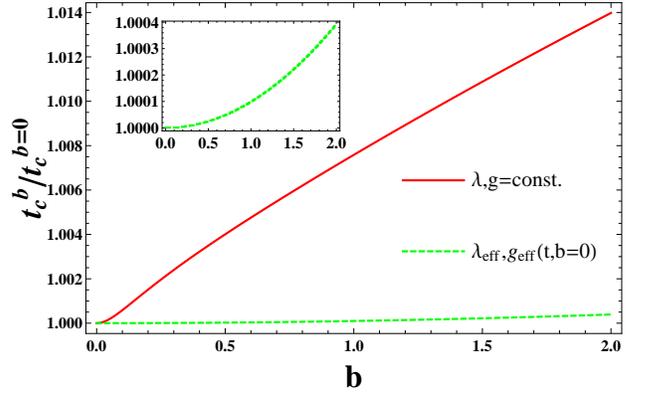}
\end{center}
\caption{Color on-line. Effect of the couplings on the critical temperature. The solid curve corresponds to the case where the couplings are taken as constants $\lambda= 0.225$ and $g=0.3$. The dashed curve
corresponds to the calculation where $g_{\mbox{\small{eff}}},\ \lambda_{\mbox{\small{eff}}}(T, qB = 0; \lambda=0.9; g=0.3)$. This last curve is also a growing function of  $b=qB/\mu^2$, as can
better be seen in the inset, though the growth is less strong
than the case computed with $\lambda$ and $g$ taken as constants.}
\label{fig6}
\end{figure}
where $P_i$ represents the incoming momentum and $m_{u,d}$ are the  $u$ and $d$-quark masses, respectively.  We emphasize that the quark masses are taken as equal and that the notation makes reference to the fact that the propagators are evaluated accounting for the different quark charges. The explicit computation is carried out in the weak field limit, with the boson propagator given by Eq.~(\ref{propnneq0}) and the fermion propagator given by~\cite{Chyi}
\begin{eqnarray}
&&S_B(K,m_f) = \frac{(m_f -\slsh{K})}{K^2+m_f^2} - i\frac{\gamma_1 \gamma_2(qB) (m_f -\slsh{K}_{\|})}{(K^2+m_f^2)^2} \nn
&+&\frac{2 (qB)^2 K_{\bot}^2}{(K^2+m_f^2)^4} \biggl[ (m_f-\slsh{K}_{\|}) + \frac{\slsh{K}_{\bot}(m_f^2+K_{\|}^2)}{K_{\bot}^2} \biggr].
\label{ferweak}
\end{eqnarray}
Using Eq.~(\ref{ferweak}) into Eq.~(\ref{general2}) and working in the high temperature limit, were we can neglect the incoming momentum in the numerator, we get up to order $B^2$
\begin{eqnarray}
L(P_i;v) &=& 2 T g^3 \sum_n\int\frac{d^3k}{(2\pi)^3} \nn 
&\times&\biggl[ \frac{1}{((P_i-K)^2+m_f^2)((P_i-K)^2+m_\pi^2)} \nn 
&-&\frac{(qB)^2}{(K^2+m_f^2)((P_i-K)^2+m_\pi^2)^3} \nn
&+& \frac{2(qB)^2 K_{\bot}^2}{((P_i-K)^2+m_f^2)((P_i-K)^2+m_\pi^2)}   \nn
&+& \frac{2(qB)^2 (K^2_{\|}+ m_f^2)}{9((P_i-K)^2+m_f^2)^4((P_i-K)^2+m_\pi^2)}       \biggr], \nn
\label{fortriangle}
\end{eqnarray}
where since we are pursuing an effective vertex correction, we have written Eq.~(\ref{fortriangle}) after averaging over the quark spins. Note that in order to handle high powers in the denominators, we can use the identity
\begin{eqnarray}
\frac{1}{(K^2+m_i^2)^{n+1}}=\frac{(-1)^n}{n!}\left(\frac{\partial}{\partial m_i^2}\right)^n\frac{1}{(K^2+m_i^2)},
\label{identity}
\end{eqnarray}
and thus, we only need to explicitly calculate the first term in Eq. (\ref{fortriangle}). The sum over Matsubara frequencies can be carried out straightforward and the result is
\begin{eqnarray}
&&\sum_n\int\frac{d^3k}{(2\pi)^3} \biggl[ \frac{1}{((P_i-K)^2+m_f^2)((P_i-K)^2+m_\pi^2)} \biggl] = \nn
&& \int{\frac{d^3k}{(2 \pi)^3}} \biggl( \frac{1-2n_f(E_\pi/T)}{E_\pi}- \frac{1-2n_f(E_f/T)}{E_f} \biggr), \nn
\label{sumMats}
\end{eqnarray}
where we have taken the infrared limit, $P_i\rightarrow 0$. $n_f$ stands for the Fermi-Dirac distribution and
\bea
E_f&=&\sqrt{\textbf{k}^2+m_f^2},\nn
E_\pi&=&\sqrt{\textbf{k}^2+m_\pi^2}.
\label{energies}
\eea
Therefore, using Eqs.~(\ref{sumMats}) and~(\ref{identity}) and performing the integration by means of the method in Ref.~\cite{Jackiw} and evaluating at $v=0$, we get the expression for the effective thermo-magnetic correction to the top diagram in column (a) of Fig.~\ref{fig4} in the infrared limit as
\begin{eqnarray}
&&L(0,0)= \frac{g^3}{ 8 \pi^2} \biggl[-2\gamma_E +1 - \ln\left( \frac{T^2 \pi^2}{2 \mu^2} \right)  \nn
&-&7\left(\frac{m_\pi^2}{8 \pi^2 T^2}\right) \zeta(3) + 31\left(\frac{m_\pi^2}{8 \pi^2 T^2}\right)^2 \zeta(5)   \nn
&+& \frac{3410}{9} \frac{(qB)^2}{ (4 \pi T)^4} \zeta(5) \biggr] ,
\label{gcorrection}
\end{eqnarray}
where we have introduced a counter term $\delta m_f^2=-1/\epsilon+\gamma_E-\ln(2\pi)$ to take care of quark-mass renormalization. Note that the quark mass does not appear since, as can be seen from Eq.~(\ref{masses}), this is proportional to $v$ and we have computed the correction for $v=0$. Therefore the correction to $g$ at one-loop order is given by
\begin{equation}
g_{\mbox{\small{eff}}}= g [1+g^2L(0,0)].
\end{equation}

In order to use a set of values for the couplings $\lambda$ and $g$ appropriate for the description of the phase transition note that the curvature of the effective potential vanishes for $v$  = 0. Since the boson thermal masses are proportional to this curvature, they also vanish at $v  = 0$. This observation provides a condition to obtain a relation between the model parameters at $T_c$ that can be supplemented with information from the physical masses of the pion and sigma in vacuum~\cite{criticalpoint}. Figure~\ref{fig5} shows the behavior of $g_{\mbox{\small{eff}}}$ as a function of $b=qB/\mu^2$ for three different values of $t= T/\mu$. Note that contrary to the behavior of $\lambda_{\mbox{\small{eff}}}$ the effective boson-quark coupling $g_{\mbox{\small{eff}}}$ is a growing function of the magnetic field strength, though the growing is rather mild.

\section{Critical temperature}\label{IV}

\begin{figure}[ht]
\begin{center}
\includegraphics[scale=0.5]{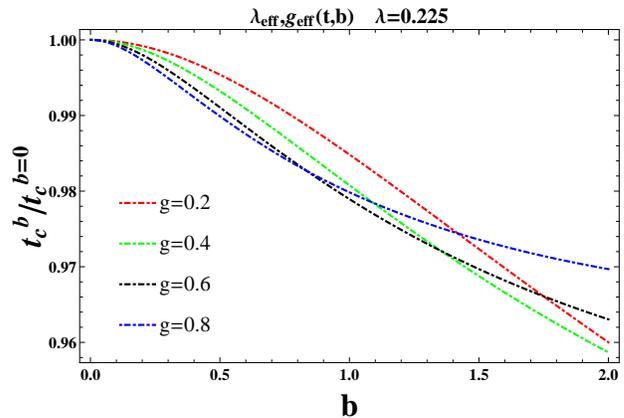}
\end{center}
\caption{Color on-line. Effect of the full thermo-magnetic dependence of couplings on the critical temperature for a fixed value of the tree level $\lambda=0.225$ and different values of the tree level $g$ as a function of $b=qB/\mu^2$. In all cases the critical temperature is a decreasing function of $b$.}
\label{fig7}
\end{figure}

\begin{figure}[ht]
\begin{center}
\includegraphics[scale=0.54]{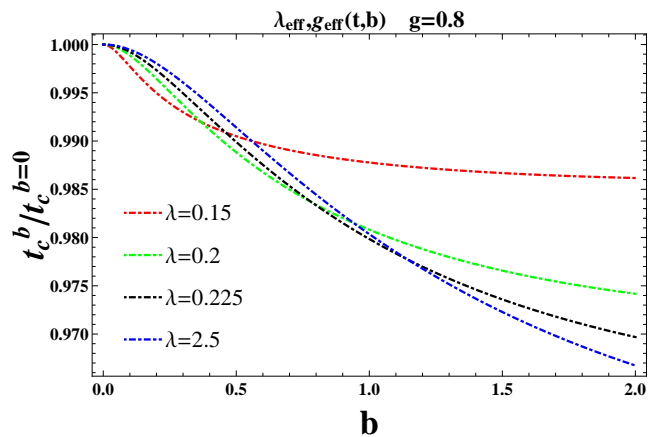}
\end{center}
\caption{Color on-line. Color on-line. Effect of the full thermo-magnetic dependence of couplings on the critical temperature for a fixed value of the tree level $g=0.8$ and different values of the tree level $\lambda$ as a function of $b=qB/\mu^2$. In all cases the critical temperature is a decreasing function of $b$.}
\label{fig8}
\end{figure}

Let us now study the effect that the thermo-magnetic corrections to the couplings have on the critical temperature. We first look at the cases where we set the couplings to their tree level values and where only thermal effects are included. Figure~\ref{fig6} shows the critical temperature in these cases, obtained from setting the second derivative of Eq.~(\ref{Veff-mid}) equal to zero at $v=0$, normalized to the critical temperature for vanishing magnetic field. Note that in both cases the critical temperature is an increasing function of the field strength though, when the thermal effects on the couplings are included, the growth is tamed. 

Figures~\ref{fig7} and~\ref{fig8} show the critical temperature for the case where we consider the full thermo-magnetic dependence of the couplings. Figure~\ref{fig7} shows the case when we set the tree-level coupling $\lambda$ to a fixed value and vary the tree-level coupling $g$.  Figure~\ref{fig8} shows the complementary case where we set the tree-level coupling $g$ to a fixed value and vary the tree-level coupling $\lambda$. Note that in all cases the critical temperature is a decreasing function of the field strength. 

\section{Summary and Conclusions} \label{conclusions}

In summary, we have shown that when including the one-loop thermo-magnetic effects for the couplings in the linear sigma model with fermions interacting with an external magnetic field, the critical temperature for the chiral transition is a decreasing function of the field strength. This behavior is a direct consequence of the decrease of the boson self-coupling with the field strength.  The effect of the fermions is marginal and the main contribution comes from the charged pions. We emphasize that the thermo-magnetic dependence of the couplings has been computed --as opposed to assumed-- within the model itself. 

In order to have a better grasp about how in this work we obtain different results from earlier works also in the context of the linear sigma model ({\it e.g.}, Ref.~\cite{Fraga:2008qn}), let us recall that for theories with spontaneous symmetry breaking the particle masses depend on the vacuum expectation value of the condensing boson, $v=v_0$ and thus when working at $T\neq 0$, the expectation value depends on $T$ and so do the particle masses.  Therefore $v_0$ plays the role of an order parameter, with the effective potential being a function of $v$ and reaching its minimum at $v=v_0$. To carry the analysis that leads to finding the change of $v_0$ with $T$ it is necessary to consider the effective potential evaluated in the full $v$-domain and consequently the dependence of the particle masses on $v$. It happens that for some $v$-values in this domain, the square of the particle masses vanish or even become negative. This is the well-known signal for the requirement to go beyond mean field and consider the resummation of the ring diagrams. The approach has been successfully implemented since the pioneering work in Ref.~\cite{Jackiw} as well as in the context of the Standard Model ({\it e.g.} Ref.~\cite{Carrington}). At $T\neq 0$ and $B=0$ inclusion of these diagrams leads to terms that cancel the offending ones. 
They also produce that for certain values of the model parameters and some temperatures the effective potential becomes an expansion containing a cubic term in the order parameter signaling that the description goes beyond the mean field approximation (see Ref.~\cite{ahmrv} for details). 

The physics involved is the proper treatment of the plasma screening effects in the infrared, encoded in these ring diagrams. At $T\neq 0$ and $B\neq 0$ the infrared problems become more severe due to the effective dimensional reduction caused by the discretization of energy levels in the direction transverse to the magnetic field. Nevertheless, it has been shown in Ref.~\cite{ahmrv} that including the ring diagrams, computed in the presence of the magnetic field, also cures these problems, thereby canceling the new offending infrared terms. In summary, the difference between our approach and previous ones within the context of the linear sigma model is that we have let the particle masses carry its dependence on $v$ and, upon inclusion of the plasma screening effects, have found the self-consistent thermomagnetic effective potential beyond mean field which is then used to perform the analysis to include thermomagnetic correction to the couplings. All together this leads to the inverse magnetic catalysis behavior.

Though the linear sigma model is not equivalent to QCD, it captures one of its low energy features, namely, chiral symmetry restoration at finite temperature, whose primary physical consequence is the change of the particle's masses as a function of the order parameter. As follows from Eqs.~(\ref{masses}), since the couplings enter as an ingredient into the boson's masses, the modification of the couplings introduces a magnetic field dependence on these masses and thus on one of the main parameters describing symmetry restoration. Since the linear sigma model used in this calculation contains no variables that make reference to the deconfinement phase transition such as the Polyakov loop, the inverse magnetic catalysis we find is determined only by parameters associated to chiral symmetry.  This finding reinforces the expectation that, since the deconfinement and chiral symmetry restoration transitions are intertwined, except for possible fine details, the description of either of the transitions in a model, which emphasizes one or the other of these aspects, must give very similar results for values of the physical parameters describing this transition, namely the critical temperature or eventually the location of the critical point.

\section*{Acknowledgments}

Support for this work has been received in part from DGAPA-UNAM under grant number PAPIIT-IN103811, CONACyT-M\'exico under grant number 128534 and FONDECYT under grant  numbers 1130056 and 1120770. R. Z. acknowledges support from CONICYT under Grant No. 21110295.

\end{document}